\documentclass[fleqn,usenatbib]{mnras}
\usepackage{newtxtext,newtxmath}
\usepackage[T1]{fontenc}
\usepackage{ae,aecompl}
\usepackage{graphicx}   
\usepackage{amsmath}    
\usepackage{amssymb}    
\title[The occurrence of SBs around WDs using K2]{The occurrence of planets and other substellar bodies around white dwarfs using K2}
\author[van Sluijs et al.]{
L. van Sluijs,$^{1}$\thanks{E-mail: vansluijs@strw.leidenuniv.nl}
V. Van Eylen,$^{1}$\thanks{E-mail: vaneylen@strw.leidenuniv.nl}
\\
$^{1}$Leiden Observatory, P.O. Box 9513, NL-2300 RA Leiden, The Netherlands\\
}
\date{Accepted XXX. Received YYY; in original form ZZZ}
\pubyear{2017}
\begin{document}
\label{firstpage}
\pagerange{\pageref{firstpage}--\pageref{lastpage}}
\maketitle
\begin{abstract}
The majority of stars both host planetary systems and evolve into a white dwarf (WD). To understand their post-main-sequence (PMS) planetary system evolution, we present a search for transiting/eclipsing planets and other Substellar Bodies (SBs) around WDs using a sample of 1148 WDs observed by K2. Using transit injections, we estimate the completeness of our search. We place constraints on the occurrence of planets and substellar bodies around white dwarfs as a function of planet radius and orbital period. For short-period ($P < 40$ days) small objects, from asteroid-sized to $1.5 \ R_{\earth}$, these are the strongest constraints known to date. We further constrain the occurrence of hot Jupiters ($< 1.5 \%$), habitable zone Earth-sized planets ($< 28 \%$), and disintegrating short-period planets ($\sim 12 \%$). We blindly recovered all previously known eclipsing objects, providing confidence in our analysis, and make all light curves publicly available.
\end{abstract}
\begin{keywords}
methods: data analysis -- occultations -- Planetary Systems -- (stars:) white dwarfs -- planets and satellites: dynamical evolution and stability 
\end{keywords}
\section{Introduction}
Most Milky Way stars are orbited by planets \citep{Cassan2012} and $\sim 95 \%$ become a WD \citep{Althaus2010}. An average WD has a radius comparable to the Earth. Therefore, an eclipsing Earth-sized planet around a WD has a $\sim10^4$ times larger transit depth than a solar-type star and would be an excellent candidate to observe faint atmospheric features or the first exomoons \citep{Loeb2013}. Furthermore, WDs host a stable ($\geq 4 \ \rm{Gyrs}$) habitable zone  \citep{Agol2011}, have a similar photosynthesis-relevant-wavelength integrated flux and DNA weighted UV dose as the Sun \citep{McCree1971, Fossati2014}, thus making them excellent candidates to host habitable planets.\\
Nevertheless, no intact planets around WDs have been found so far \citep{Farihi2005, Mullally2006, Mullally2008, Kilic2009, Drake2010, Faedi2011, Fulton2014, Vanderburg2015, Xu2015, Sandhaus2016}. This null detection is in agreement with theoretical simulations of PMS planetary system evolution: the inner planets are destroyed due to tides during the stellar AGB-phase \citep{Villaver2009, Kunitomo2011, Mustill2012, Villaver2014}. The outer planets planets expand their orbits due to stellar mass loss during the AGB-phase \citep[see. e.g.][]{Veras2016a, Veras2016b}.\\
However, several indirect observations of remnant planetary systems exist. About 25-50$\%$ of all WDs have metals in their photosphere \citep{Zuckerman2003, Zuckerman2010, Koester2014}. Typical metal sinking time scales are of the order of days to Myrs \citep{Wyatt2014}, much shorter than the typical WD age, therefore these WDs are called metal polluted. Additionally, about $4.6 \%$ of all of the WDs host a debris disk \citep{Wyatt2014}. The amount of accreted material matches asteroid-sized objects \citep{Xu2011,Girven2012, 2010MNRAS.404.2123F} and the compostion matches Solar System meteorites and bulk Earth to zeroth order \citep{2014AREPS..42...45J}. This suggests metal pollution and debris disks originate from material of remnant planetary systems.\\
Thus, a missing link must exist between these observations and planetary system evolution theory. The most common explanation is scattering of asteroids by unseen planet towards the WD \citep{Bonsor2011, Bonsor2015, Debes2012, Dong2010, Frewen2014, Payne2016}. This is supported by simulations showing that AGB-phase stellar mass-loss can cause planets to be perturbed inwards during the WD-phase \citep{Veras2014, Payne2016}. The recent discovery of a disintegrating planet around a WD \citep{Vanderburg2015} provides additional evidence favoring this scenario.\\
This discovery and the previous arguments suggest short-period SBs around WDs might exist. An opportunity presents itself with the K2 mission \citep{Howell2014}, which observed photometry of a significant number of WDs. Here, we present a search for new transiting/eclipsing SBs around 1148 WDs observed in the first 13 campaigns of K2. We also determine the recovery rate of transiting objects by injecting artificial transit signals and calculating how many injections were successfully recovered by the transit detection algorithm. Combining the number of transiting/eclipsing SBs and the K2 detection probability allows us to constrain the occurrence of planets and other SBs around these stars. In Section~\ref{sec:data}, we provide an overview of the K2 WD data set. We explain the adopted methods in Section~\ref{sec:methods}. The results are presented in Section~\ref{sec:results} and discussed in Section~\ref{sec:discussion}. We draw conclusions in Section~\ref{sec:conclusions}. 
\section{Data}
\label{sec:data}
The dataset used in this work has been collected by the Kepler satellite during campaigns 1-13 of the K2 mission. During these campaigns, many WDs have been observed for various reasons: to detect transiting SBs, to study eclipsing binary stars, to examine magnetic fields, to study WD pulsations and to search for rotational modulation of WDs. Combined, the initial sample of WDs consists of 1610 targets. An overview of included programs is given in Table~\ref{tab:programs}.\\
Amid these programs a wide diversity of target selection methods exists: some programs only selected spectroscopically confirmed WDs where others used photometric colour and/or reduced proper motion cuts. The latter leads to contamination of the sample by non-WDs with similar photometric and proper motion properties. Secondly, some targets have a main-sequence (MS) companion that contributes a significant fraction of the observed flux. 
Therefore, the total sample has been cross-correlated with a list of high-probability and confirmed white dwarfs \citep{hermes2017}\footnote{An updated list of {\em K2} confirmed and high-probability candidate WDs can be found at: \url{http://www.k2wd.org} (J.~J.~Hermes 2017, private communication).}.
Lastly, there are a dozen duplicate objects within the sample. In total we counted 364 non-confirmed WDs, 90 composite WDs and 39 duplicates within the initial sample of 1610 WDs. Combined, this reduces the total sample to 1148 WDs.\\
\begin{table}
        \caption{All included programs constituting the total sample of 1610 targets.}
    \label{tab:programs}
        \centering
    \begin{tabular}{ l r }
    \hline
    {Campaign} & {Programs} \\
    \hline
    1 & GO0122, GO0110, GO0071, GO0010, GO0001, \\
     & GO1071, GO1048, GO1012, GO1007, GO1004 \\
    2 & GO02111, GO2087 \\
    3 & GO3116, GO3111, GO3087, GO3005 \\
    4 & GO4073, GO4043, GO4041, GO4017, GO4003, \\
     & GO4001 \\
    5 & GO5073, GO5043, GO5041, GO5017, GO5003, \\
     & GO5001 \\
    6 & GO6083, GO6063, GO6050, GO6045, GO6003, \\
    & GO6001 \\
    7 & GO7063, GO7050, GO7045, GO7003, GO7001 \\ 
    8 & GO8048, GO8019, GO818, GO8011, GO8006 \\
    9 & - \\
    
    10 & GO10006, GO10902, GO10076, GO10048, \\ 
    & GO10019, GO10018, GO10011, GO10006 \\
    11 & GO11040 \\
    12 & GO12901, GO12040, GO12037, GO12027 \\
    & GO12007 \\
    13 & GO13040, GO13037, GO13027, GO13007 \\
    \hline
    \end{tabular}
\end{table}
All data files (target pixel files) consist of a time series of images of the target. All initial 1610 targets have been observed with long cadence (30 minutes exposure time). Among these, 374 targets have also been observed with short cadence (1 minute exposure time) of which 309 targets are confirmed, non-composite WDs. Since an Earth-sized planet around an average WD has a transit duration of $\sim$2 minutes \citep{Agol2011}, long cadence transits are diluted. Therefore, smaller transiting/eclipsing objects can potentially be detected in the short cadence data.
\section{Methods}
\label{sec:methods}
Transiting/eclipsing Stellar and Substellar Bodies (SSBs) cause a periodic drop of the total stellar flux. This research aims to detects these signals and to evaluate their nature. The initial total sample of 1148 WDs was analysed for this purpose. The amount of SSBs in the data set constrains their occurrence. In Section~\ref{sec:dataredlightcurves}, we describe how the pixel target files are converted into light curves. In Section~\ref{sec:bls}, we discuss the transit/eclipse detection algorithm, and in Section~\ref{sec:transitinjection}, we explain the transit injection procedure, which is necessary to determine the recovery rate of K2 transits. Finally, in Section~\ref{sec:occurrence} we describe how to combine the transit search with the recovery rate to constrain the transit occurrence.
\subsection{Conversion of raw data into light curves}
\label{sec:dataredlightcurves}
To have full control over our analysis and injection pipeline, we start from the raw data, in the form of target pixel files, which can be downloaded from the MAST archive\footnote{\url{https://archive.stsci.edu/k2/}}. We then convert the target pixel files into light curves using aperture photometry, and correct for systematic-errors, such as those caused by the pointing jitter in the two-wheel K2 satellite \citep{Howell2014}. Our pipeline to do this is largely based upon the publicly available pipeline\footnote{\url{https://github.com/vincentvaneylen/k2photometry}} by \cite{Vaneylen2016} and to a lesser extent on other previous research \citep[see e.g.][]{Lund2015, Vanderburg2015, Sanchis-Ojeda2015}. We optimized our pipeline for detection of objects around WDs, and describe its main characteristics here\footnote{The pipeline is publicly available at \url{https://github.com/lennartvansluijs/WD-pipeline-K2}.}:
\begin{enumerate}
\item Initial flagging: A fraction of the data is flagged using the {\sc QUALITY} tag\footnote{More information on the {\sc QUALITY} tag can be found in the Kepler archive manual \url{http://archive.stsci.edu/kepler/manuals/archive_manual.pdf}} available for every pixel target file. This describes when special spacecraft events occurred or when the pipeline flagged a phenomenon. Most of these cannot be triggered by a transit/eclipse signal and are removed. Non-removed tags are: 0 (no event noted), 512 (unused bit by Kepler), 2048 (impulsive outlier before co-trending) and 136426 (local detector electronics parity errors). Trial-and-error found the last tag (2048) to flag in-transit data points and therefore is kept. Tag 136426 mostly affects campaigns 2 and 11, but for these campaigns K2 data release notes state the data quality is not adversely affected.
\item Aperture photometry: Initially, for the long cadence, irregular apertures are created, starting from the brightest pixel in the concatenated pixel target file and by subsequently adding the brightest neighbouring pixel to the previous aperture, until a maximum of 25 pixels. Using all 25 apertures, the data is reduced in accordance to the description below and the aperture that results in the minimal standard deviation in the flux of the final light curve is chosen as the preferred light curve. For the short cadence observations, the best aperture from the long cadence observations is chosen, because creating  25 apertures for these data is time consuming. For campaigns $> 2$ background subtraction is performed already by the K2 data reduction pipeline. For earlier campaigns, the background flux is estimated for all images in the pixel target file by the median flux of all pixels outside of the best aperture, and this value is subtracted. Finally, we add all flux within the best aperture for the background-removed images to generate the raw light curve. Any 'inf', 'NaN' or 'zero' values are flagged and the light curve is divided by the median.
\item Thruster event removal: Every $\sim 6 \ \rm{h}$, the K2 satellite fires thrusters to maintain its pointing. These events lead to flux outliers and must be removed. To do this, the center-of-flux is calculated for all images in the pixel target file. The differences between subsequent center-of-fluxes are calculated. For the long cadence two data points around differences larger than 3$\sigma$ are removed. For the short cadence, where thruster events affect multiple data points, five data points around differences larger than 3.5$\sigma$ are removed.
\item Self-flat-fielding: 
To separate long-term $\geq$24 h photometric variations from Kepler's slow drift every $\sim 6.5 \ \rm{h}$, self-flat-fielding is performed. The data is sliced into parts of $\sim 24 \ \rm{h}$ and binned. A B-spline is fitted to the binned data, with 3$\sigma$-outliers flagged to ensure a good fit. Division by the best fit decorrelates for the long-term variations.
\item Drift correction: Firstly, the data set is sliced into 11 parts ($\sim 6.5 \ \rm{days}$ per part), a timescale for which the satellite's drift has been observed to be quite stable along its roll axis. Secondly, for every slice a principal components analysis is performed to change to a coordinate system where the star moves only along the main direction. A fourth degree polynomial as a function of the moving coordinate is fitted to every slice, with 2$\sigma$-outliers flagged to ensure a good fit. The fluxes are decorrelated against the moving coordinate by division of the best fit. The processes of self-flat-fielding and drift correction are iterated three times to ensure both converted.
\end{enumerate}
All 1148 long cadence and 309 short cadence systematic-error-corrected light curves have been made publicly available\footnote{The reduced data is publicly available at \url{https://github.com/lennartvansluijs/WD-pipeline-K2}.}.
\subsection{The transit detection algorithm}
\label{sec:bls}
After converting the raw data into light curves, transit/eclipse signals must be found. For the long cadence, transit/eclipse dilution causes these signals to become very box-like. Therefore, an algorithm optimized to search for box-like signals is used: the Box-Least-Square (BLS) algorithm \citep{Kovacs2002}. We used a Python implementation of this algorithm\footnote{\url{https://github.com/dfm/python-bls}, by Ruth Angus and Dan Foreman-Mackey.}. The BLS algorithm folds a light curve for $n_P$ periods within the range $(P_{\rm{min}}, P_{\rm{max}})$, bins the data into $n_B$ bins, fits boxes with relative widths within $(q_{\rm{min}}$, $q_{\rm{max}})$ and calculates the signal-residue $SR(P)$, the BLS-spectrum. Here $n_{P} = 10,000$ and $n_B = 300$ is used, which have sufficient sampling and short computational time. We choose $P_{\rm{min}} = 1 \ \rm{h}$, well-below the Roche radius for giants $\sim 5 \ \rm{h}$ \citep{Fulton2014}. $P_{\rm{max}}$ is set to half the total campaign length, i.e.\ approximately 40 days. Finally, we choose $q_{\rm{min}} = 10^{-4}$ and $q_{\rm{max}} = 0.05$, corresponding to a 5 h Earth ($\sim$minimal duration) and a 80 d Jupiter ($\sim$maximal duration) around a WD.\\
From the BLS-spectrum, three best candidate periods are selected. The first candidate corresponds to the BLS-spectrum maximum. The second and third candidate correspond to the second and third BLS-spectrum maximums, under the extra condition that they are no harmonics of the best candidate period or each other. Harmonics $H$ are defined as periods $P$ for which $\{ H: \frac{P}{n}, \frac{P}{n-1}, ..., P, ..., (n-1)P, nP \in (H - H \Delta H, H + H \Delta H) \}$. Here $\Delta H = 2 \%$ and $n = 11$ are used to ensure candidate periods are not degenerate, but not to exclude other periods.\\
An overview figure is generated containing the unfolded systematic-error reduced light curve, the BLS-spectrum and three folded systematic-error reduced light curves. All 1148 targets were manually inspected for transit/eclipse signals. Those with large signal-residues for the candidate periods were inspected with extra care, since leftover correlation due to Kepler's pointing jitter can also induce large signal-residues. Furthermore, all targets were inspected for contaminant stars, especially for those where an eclipse/transit signal is detected: these targets were additionally reduced using manually defined apertures encircling only the target star, to ensure the signal does not originate from a nearby field star.
\subsection{Transit injection}
\label{sec:transitinjection}
The observed occurrence equals the number of SSBs $n$ over the total sample size $N$. However, this is not the true occurrence: not all SSBs cause an eclipse/transit or are recovered by Kepler. The true occurrence equals $n/N'$ where $N'$ is the effective sample size defined as
\begin{equation}
        {N'(P, R_{\rm{p}}) = p_{\rm{tra}}(P, R_{\rm{p}}) \times p_{\rm{det}}(P, R_{\rm{p}}) \times N},
        \label{eq:effsamplesize}
\end{equation}
where $p_{\rm{tra}}$ is the transit probability, $p_{\rm{det}}$ the detection efficiency (sometimes referred to as coverage or completeness), $P$ the orbital period and $R_{\rm{p}}$ the SB radius. The transit probability is fully constrained by the geometry and equals
\begin{equation}
        {p_{\rm{tra}} = \frac{R_{\rm{p}} + R_{\rm{WD}}}{a}},
        \label{eq:transitprobability}
\end{equation}
where $R_{\rm{WD}}$ is the WD radius and $a$ the orbital semi-major axis. $a$ is determined from Kepler's third law assuming a WD mass of $0.6 \ \rm{M_{\rm{\sun}}}$ and radius $0.012 \ \rm{R_{\rm{\sun}}}$ \citep[following][]{Fulton2014}.\\
The detection efficiency depends on both the data quality and the data reduction and is therefore hard to theorize. It can be empirically determined using transit injection. This method injects an artificial transit/eclipse signal into the raw light curve for a given period and SB radius. We use the Mandel and Agol model \citep{Mandel2002} and inject transit signals just before systematic-error-correction (in between step (ii) and (iii) of Section~\ref{sec:dataredlightcurves}). If the right period or an harmonic is recovered, one assumes the same would hold for a real transit/eclipse signal.\\
Firstly, the ($P$, $R_p$)-plane is divided into 35 tiles by a grid of 5 logarithmically-equally spaced periods between $1 \rm{h} - 40 \rm{d}$ and 7 logarithmically-equally spaced radii between 0.125 - 16 $R_{\earth}$. For all injections, a WD of $0.6 \ \rm{M_{\rm{\sun}}}$ and radius $0.012 \ \rm{R_{\rm{\sun}}}$ is assumed again. The ellipticity is assumed to be zero, since strong tides are theorized to circularize SB orbits \citep{Nordhaus2012, Mustill2013}. The period and planet radius per injection are picked randomly within the tile range. The impact parameter $b$ is chosen uniformly within the transit range $(0, b_{\rm{max}})$ where
\begin{equation}
        {b_{\rm{max}} = \frac{R_{\rm{WD}} + R_p}{R_{\rm{WD}}}}.
        \label{eq:bmax}
\end{equation}
For each injection, we pick a random data file from our sample, and we choose a random transit epoch. Many injections per tile must be performed for the detection efficiency to converge to a mean value: 250 injections/square are performed for the long cadence and 110 injections/square for the short cadence, where the computation time is much longer. To ensure the detection efficiency truly represents the K2 detection efficiency for a mixed sample of authentic WDs, injections have been performed only in the sample of 1148 confirmed WDs of which 309 were also observed in short cadence. The detection efficiency equals the recovered injections over the total injections per tile.
\subsection{Constraining the occurrence}
\label{sec:occurrence}
The SSB detection or null-detection combined with the effective sample size constrains the occurrence. A method similar to \cite{Faedi2011} is used here.\\
Define the fraction of WDs with a SB of radius $R_{\rm{p}})$, at period $P$ as $f(P, R_{\rm{p}})$. The probability of finding $n$ SSBs in a sample of $N$ WDs, assuming all SSBs are detectable, is given by a Binominal probability distribution \citep[see e.g.][]{Burgasser2002,McCarthy2004}
\begin{equation}
        {p(n; N, f) = \binom{N}{n} f^{n} (1-f)^{N-n} }.
        \label{eq:binom}
\end{equation}
Not all planets are detectable and $N$ must be replaced by the effective sample size $N'$. In the general case of non-zero detections, the occurrence is constrained to the interval $(f_{\rm{min}}, f_{\rm{max}})$ at a confidence level $C$ when satisfying
\begin{equation}
{\int_{f_{\rm{min}}}^{f_{\rm{max}}} (1+N')^{-1} \times p(n; N', f) = C},
        \label{eq:Cinterval}
\end{equation}
where $(1+N')^{-1}$ is a normalization factor. In the special case of a null detection $n = 0$ and $f_{\rm{min}} = 0$
\begin{equation}
        {\int_{0}^{f_{\rm{max}}} (1+N')^{-1}  \times p(0; N', f) = 1 - (1-f_{\rm{max}})^{N'+1} = C},
\end{equation}
which can be solved for $f_{\rm{max}}$,
\begin{equation}
        {f_{\rm{max}} = 1 - (1-C)^{\frac{1}{N'+1}}},
        \label{eq:fmax}
\end{equation}
calculating the maximum occurrence $f_{\rm{max}}$ at a confidence level $C$ for a null detection.\\
Since the short cadence data has a better detection efficiency than the long cadence, the best effective sample size is the combined long cadence and short cadence effective sample size defined as
\begin{equation}
        {N'_{\rm{com}} = N_{\rm{SC}}' + (N_{\rm{LC}} - N_{\rm{SC}}) \times p_{\rm{tra, LC}} \times p_{\rm{det, LC}}}.
        \label{eq:effsamplesizeLCSC}
\end{equation}
This can be plugged into Equations~\ref{eq:Cinterval} or~\ref{eq:fmax} to calculate the occurrence constraints similarly.
\section{Results}
\label{sec:results}
The results are composed of three main components: Firstly, Section~\ref{sec:SSBs} gives the final list of detected eclipsing/transiting SSBs. Secondly, Section~\ref{sec:transitinjectionresults} discusses the transit injection results. Finally, Section~\ref{sec:occurrenceconstraints} presents the occurrence constraints.\\
\subsection{Detected eclipsing/transiting SSBs}
\label{sec:SSBs}
\begin{figure*}
        \centering
    \includegraphics[width = 2\columnwidth]{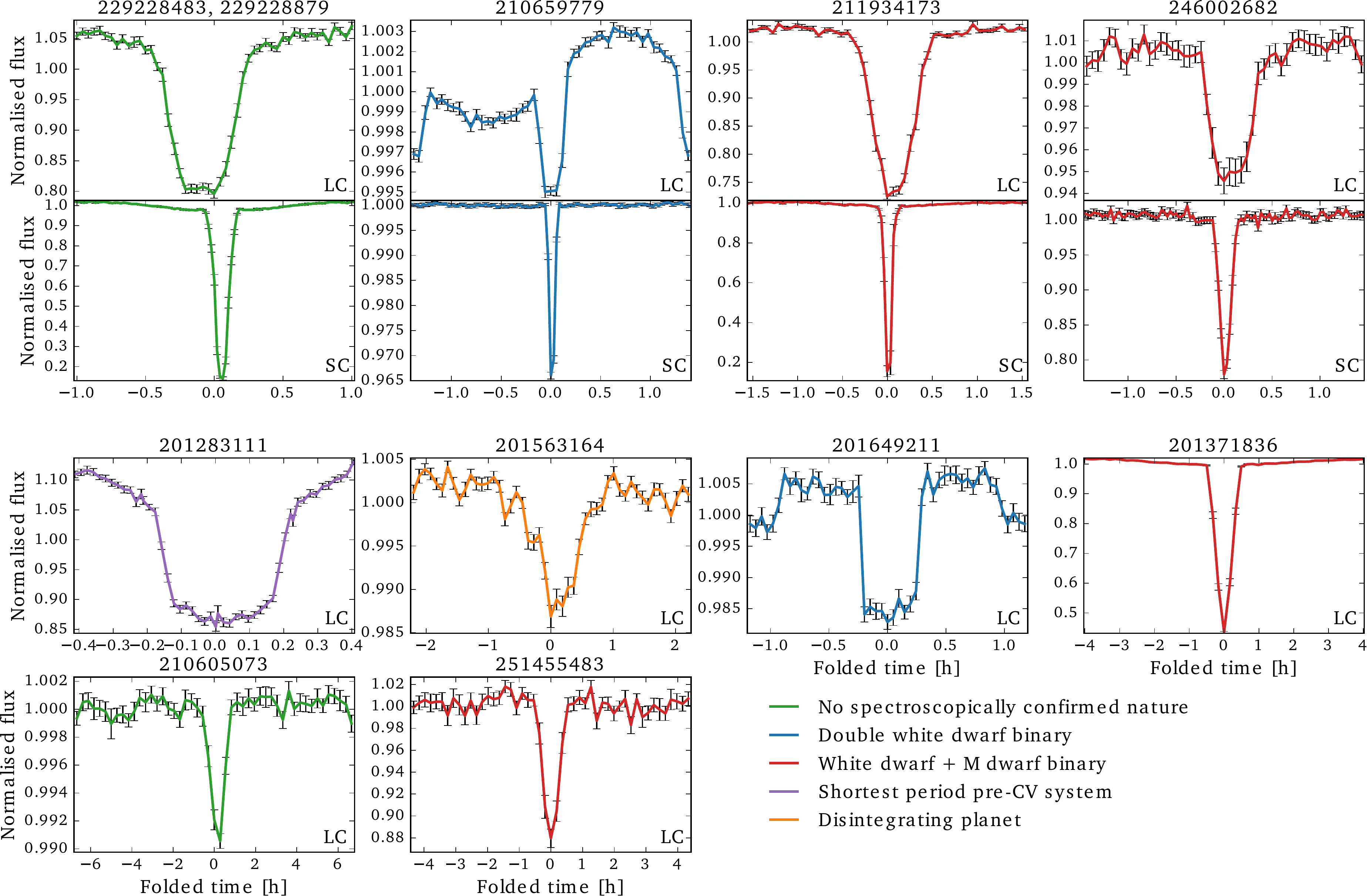}
    \caption{An overview of the detected eclipsing/transiting SSBs. All light curves are folded for the best period found by the BLS-algorithm. The data is binned after folding using 50 bins for the long cadence ('LC') data and 100 bins for the short cadence ('SC'). The error bars indicate the standard deviation per bin. The colors indicate the natures of the systems.}
    \label{fig:detectedssbs}
\end{figure*}

In total, 10 eclipsing/transiting SSBs have been detected, of which 4 were also observed in short cadence. An overview of the folded light curves is shown in Figure~\ref{fig:detectedssbs}. Many of these eclipsing/transiting objects were already known pre-K2 and selected to be re-observed by K2 to study their properties in more detail. All others were discovered earlier in K2 campaigns. All previously known eclipsing/transiting SSBs detectable in the K2 data sample have been blindly re-obtained, adding confidence to the null detection of new SBs.\\
The disintegrating planet discovered by \cite{Vanderburg2015}, EPIC 201563164, is re-obtained. EPIC 201649211 and 210659779 are both spectroscopically confirmed WD-WD binary systems \citep[see e.g.][]{Girven2012, Silvotti2012} and EPIC 201371836, 211934173, 246002682 and 251455483 are all spectroscopically confirmed M dwarf-WD binary systems \citep[see e.g.][]{Silvestri2006, Heller2009, Rebassa-Mansergas2009}. Recently, EPIC 201283111 was tentatively concluded to be the shortest-period pre-Cataclysmic-Variable star known today \citep{Rappaport2017, Parsons2017}. \cite{Parsons2017} additionally present EPIC 248368963, likely a WD-brown dwarf binary, but due to dilution of the eclipse there is only slight evidence of an eclipse in the K2 data. Such objects are not recovered by our transit injection algorithm and therefore this object has not been included in our list here. EPIC 210605073 was a proposed WD candidate, but a recent photometric analysis suggests the star is (contaminated with) an F0 companion \citep{Adams}. The companion may not be a planet: it likely has a significant volatile component, Roche-lobe overflow and/or photo-evaporative mass loss \citep{Adams}. The same presumably applies to the previously observed companion around EPIC 229228483 \citep{Kleinman2004}.
\subsection{Transit injection results}
\label{sec:transitinjectionresults}
\begin{figure}
        \centering
    \includegraphics[width = \columnwidth]{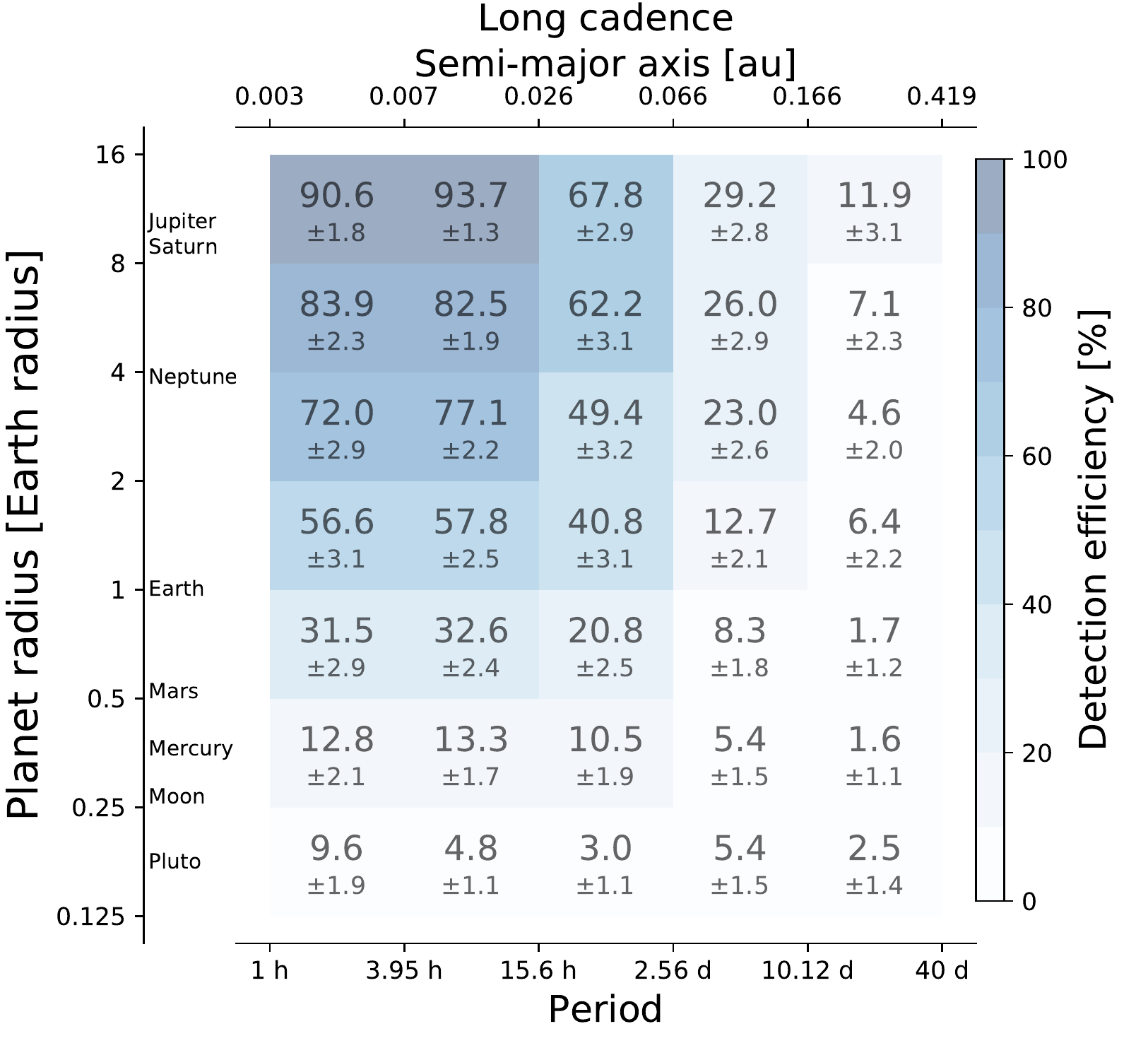}
    \includegraphics[width = \columnwidth]{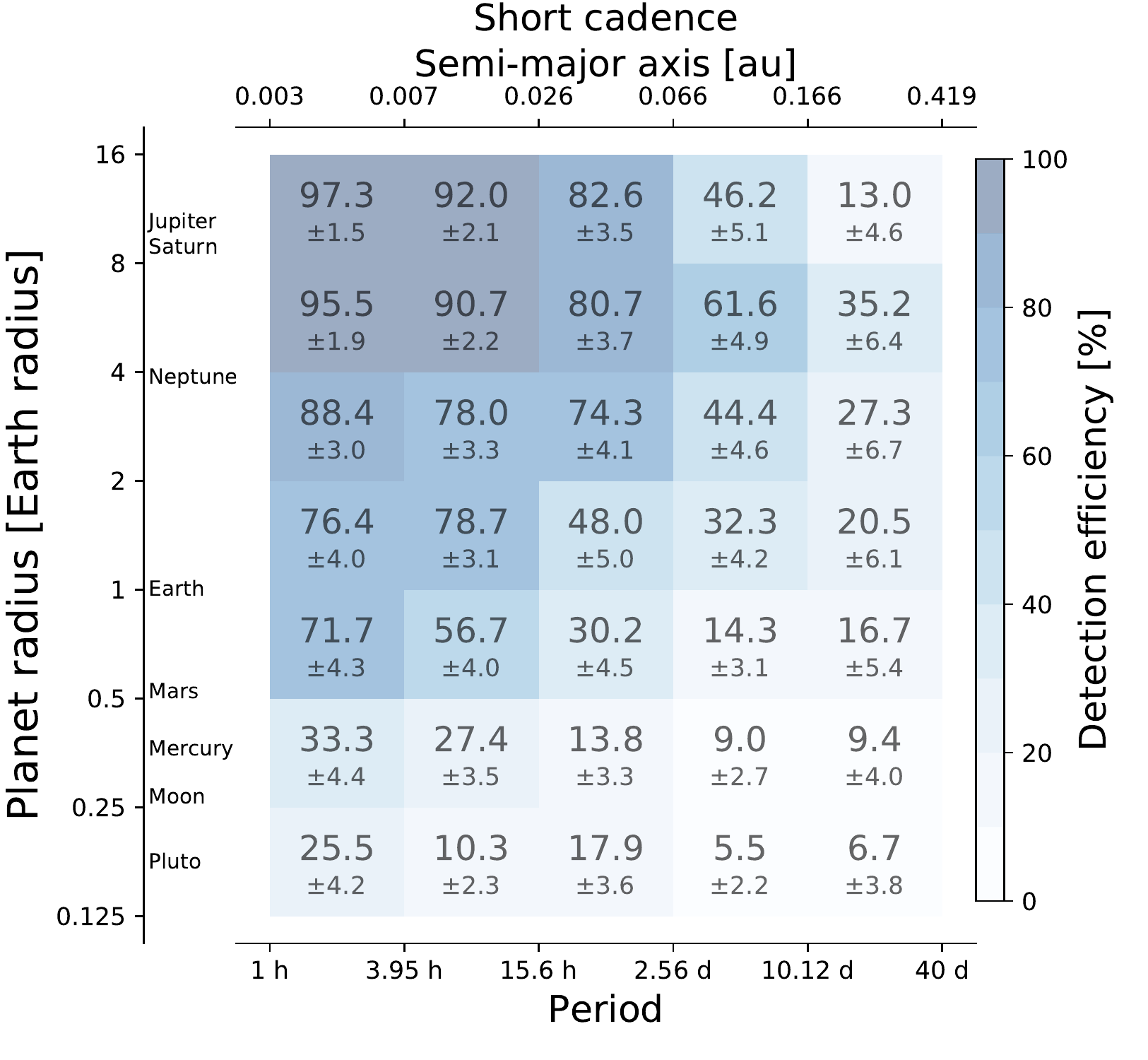}
    \caption{The top panel shows the detection efficiency for the long cadence and the bottom panel for the short cadence. A WD with a mass of $0.6 \ M_{\odot}$ and a radius of $0.012 \ R_{\odot}$ is implicitly assumed. The error bars have been calculated by bootstrap re-sampling the injections. Periods and planet radii are logarithmically-equally-spaced. Radii of some solar-system objects are indicated as a comparison.}
    \label{fig:coverage}
\end{figure}

The detection efficiency for the long cadence and short cadence are shown in Figure~\ref{fig:coverage}.
The error bars indicate the error due to a finite amount of injections and are calculated with bootstrap re-sampling. Bootstrap re-sampling draws $n$ observations with replacement from the original series to create a bootstrap re-sampled series $\{P_i^*\}_j$ used to determine the re-sampled detection efficiency $p_{\rm{det},j}^*$. This process is iterated $B = 10,000$ times and outputs a bootstrap re-sampled detection efficiency distribution $p_{\rm{det},1}^*, ..., p_{\rm{det},B}^*$. The standard deviation $\sigma$ of this distribution measures the error due to a finite amount of injections. The short cadence errors are larger due to less injections.\\
Large periods have fewer observed transits and smaller SBs have smaller transit depths, causing a lower detection probability. For the long cadence, objects bigger than $1 \ R_{\oplus}$ for periods $< 16 \ \rm{h}$ are more often recovered than not recovered. Smaller SBs can be detected in the short cadence data. For the short cadence, objects bigger than $0.5 \ R_{\oplus}$ for periods $< 16 \ \rm{h}$ are more often recovered than not recovered. Outside these ranges recovery rates are much lower, but still some injections are recovered. This reflects the diversity of the data quality and data reduction in the sample of WDs observed by K2. Overall, Figure~\ref{fig:coverage} shows that K2 is an excellent survey when searching for transiting SSBs around WDs, with the potential of detecting (sub-)Earth-sized planets around WDs.
\subsection{The occurrence constraints}
\label{sec:occurrenceconstraints}
\begin{figure}
        \centering
    \includegraphics[width=\columnwidth]{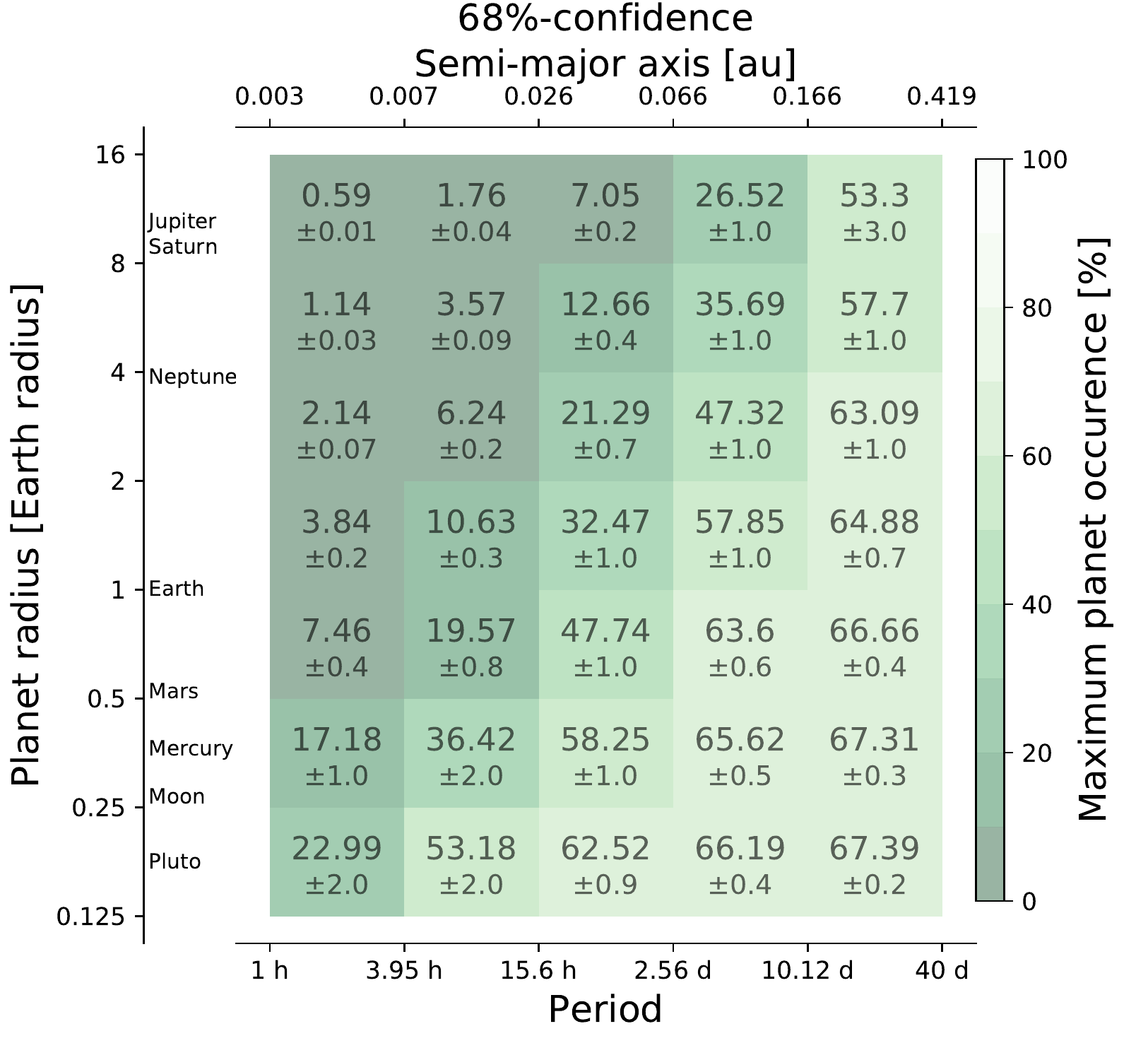}
    \includegraphics[width=\columnwidth]{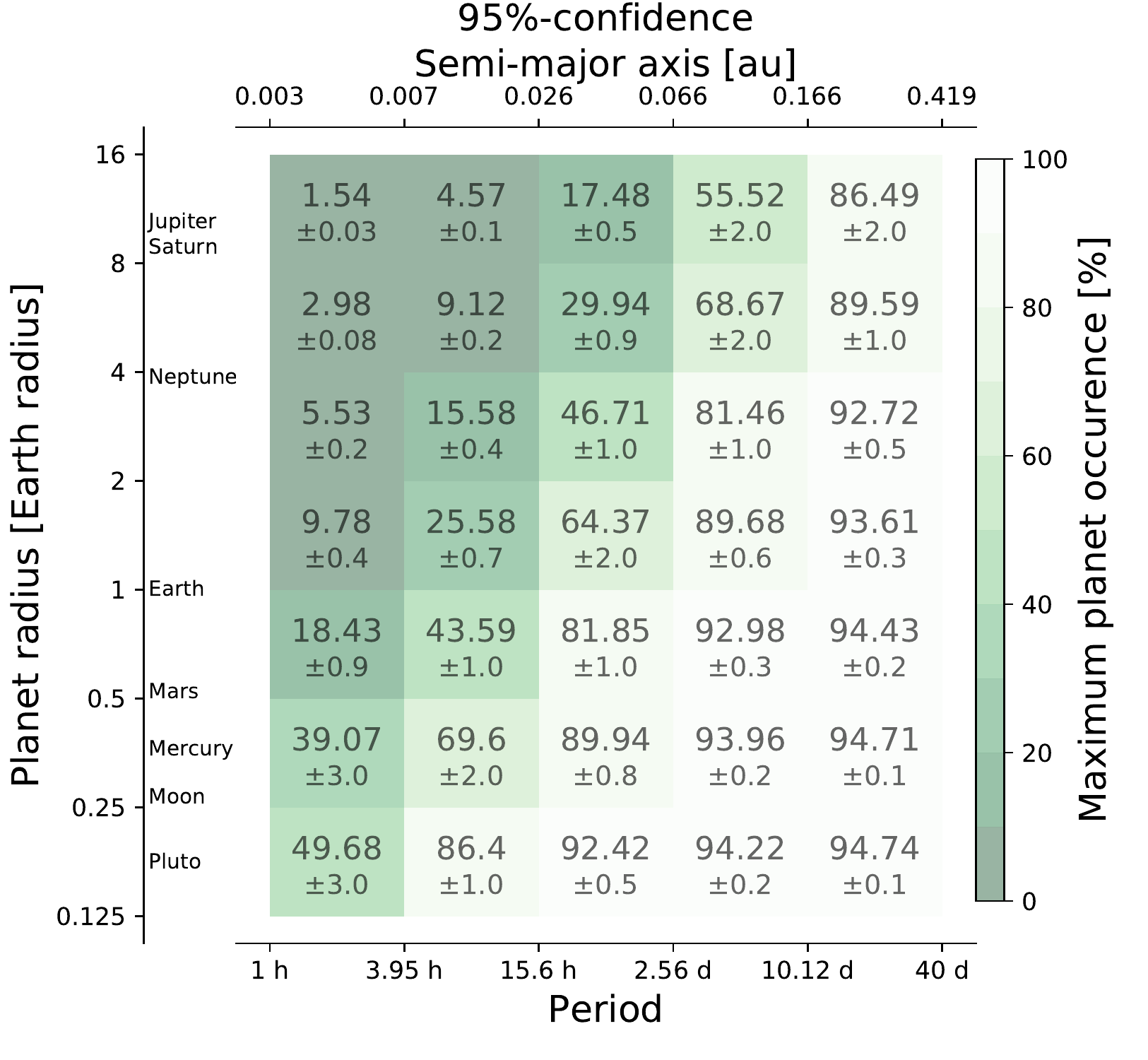}
        \caption{The maximum non-disintegrating SB occurrence around WDs for the combined long and short cadence. The top figure, respectively bottom figure, show the 68$\%$- interval and a 95$\%$-confidence interval. A WD with a mass of $0.6 \ M_{\odot}$ and a radius of $0.012 \ R_{\odot}$ is implicitly assumed. The error bars have been calculated by propagating the finite injection errors of the detection probability and average transit probability. Periods and planet radii are logarithmically-equally-spaced. Radii of some solar-system objects are indicated as a comparison.}
    \label{fig:occurrenceLCandSC}
\end{figure}

\begin{figure}
        \centering
    \includegraphics[width=\columnwidth]{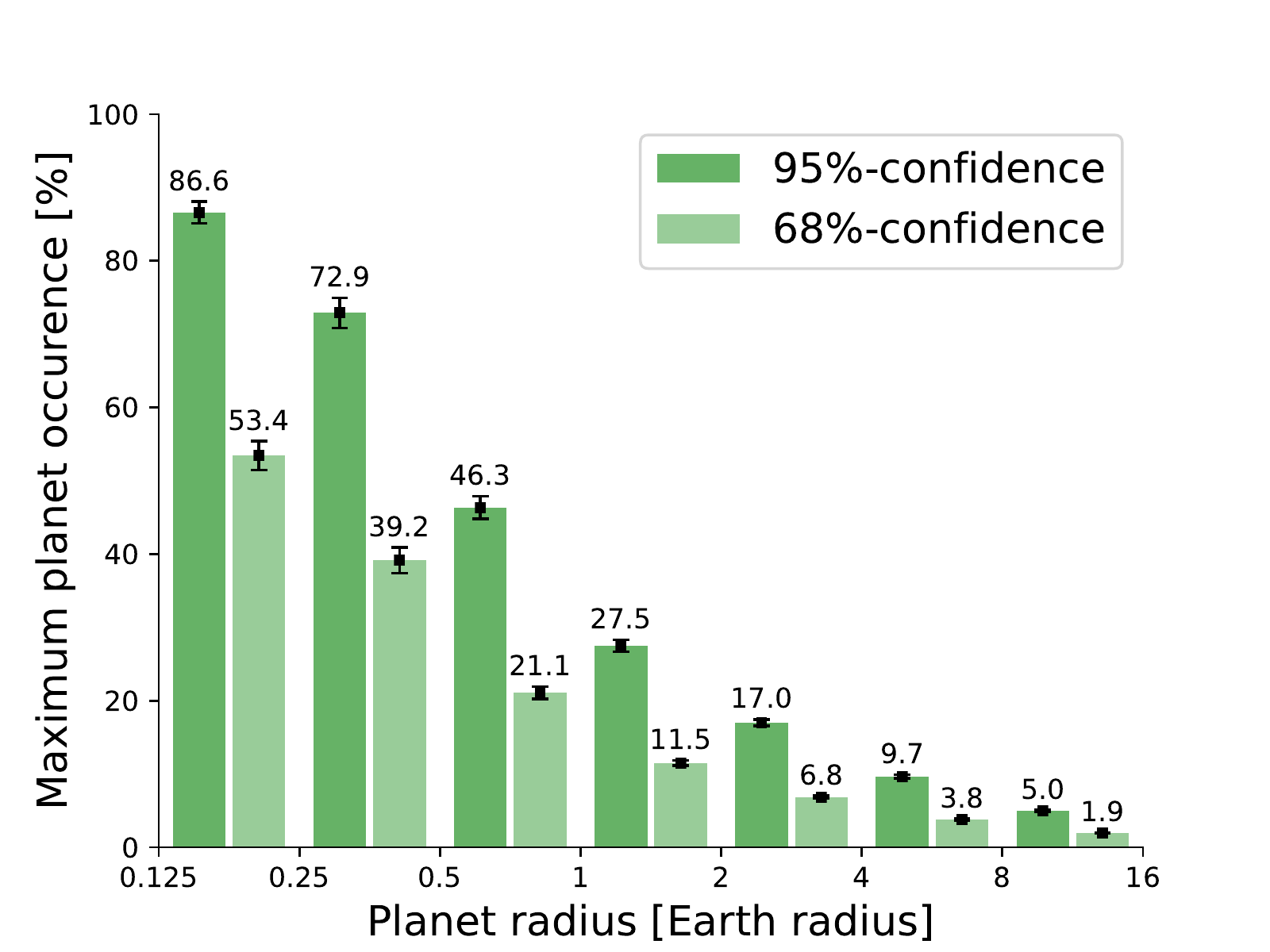}
        \caption{The maximum non-disintegrating SB occurrence around WDs in the CHZ for the combined long and short cadence for a 95$\%$- and 68$\%$-confidence interval. A WD with a mass of $0.6 \ M_{\odot}$ and a radius of $0.012 \ R_{\odot}$ is implicitly assumed. The error bars have been calculated by propagating the finite injection errors of the detection probability and average transit probability. Planet radii are logarithmically-equally-spaced.}
    \label{fig:HZconstraints}
\end{figure}

No non-disintegrating planets and other SBs have been found, therefore Equation~\ref{eq:fmax} is used to constrain their occurrence. Figure~\ref{fig:occurrenceLCandSC} presents the occurrence constraints, using the combined long and short cadence effective sample size.
The error bars are the propagated transit probability and detection probability errors due to the finite transit injections. Using the basic rule of error propagation $f(x_1, ..., x_N) = \sqrt{\sum_{i=1}^N ({\partial f} \mathbin{/} {\partial x_i}) \Delta x_i}$ the error on the effective sample size equals
\begin{equation}
        {\sigma_{\rm{N'}} = \sqrt{ \Big( p_{\rm{det}} \times N \times \sigma_{p_{\rm{tra}}} \Big)^2 + \Big( p_{\rm{tra}} \times N \times \sigma_{p_{\rm{det}}} \Big)^2  } }.
        \label{eq:sigmaeffsample}
\end{equation}
Combining the long and short cadence (see Equation~\ref{eq:sigmaeffsample}) ${\sigma_{N'}^2 =  \sigma_{N'_{\rm{LC}}-N'_{\rm{SC}}}^2 + \sigma_{N'_{\rm{SC}}}^2}$ and propagating the error to the maximum occurrence equals
\begin{equation}
        {\sigma_{f_{\rm{max}}} = \Big{|} \frac{\partial f_{\rm{max}}}{\partial N'} \sigma_{N'} \Big{|} = \Big{|} \frac{ \ln{(1-C)} (1-C)^{1/N'}}{N'^2} \sigma_{N'} \Big{|}}.
\end{equation}
These errors are much smaller than the differences between the two confidence intervals. Equivalently, the finite WD sample size dominantly constrains the maximum occurrence and not the transit injections. Since the transit injection errors are negligible, the errors are dropped from this point. The rest of this work discusses the 95$\%$-confidence interval limits, but the 68$\%$-confidence interval limits are mentioned within brackets.\\
We further investigate the occurrence rate of planets in the habitable zone. \cite{Agol2011} defined the Continuous Habitable Zone (CHZ), i.e.\ the region around an average WD where liquid water is sustainable for $> 4 \ \rm{Gyrs}$ around a WD with mass $0.6 \ M_{\odot}$ and radius $0.01 \ R_{\odot}$, close to our radius of $0.012 \ R_{\odot}$, as having $P = 4-32 \ \rm{d}$, or a semi-major axis $a = 0.005-0.02$~au. We calculate occurrence constraints by selecting only injections within this period range and show the result in Figure~\ref{fig:HZconstraints}.\\
In contrast to the non-disintegrating SB null detection, one detected disintegrating planet exists. Assuming all disintegrating planets have similar transit signals, occurrence constraints can be derived, as done previously by \cite{Vanderburg2015}. The transit probability is estimated $p_{\rm{tra}} \simeq 0.02$ \citep{Vanderburg2015}. The detection probability is assumed equal to that of a non-disintegrating body transit with a similar transit depth: $p_{\rm{det, LC}} \simeq 30 \%$ for the long cadence and $p_{\rm{det, SC}} \simeq 57 \%$ for the short cadence. Using these numbers to calculate the effective sample size, a disintegrating planet occurrence of $\sim 12 \%$ is found. To calculate the confidence interval in the general case, Equation~\ref{eq:fmax} is not valid. Instead a Jeffreys interval is used, because its property of being equally-tailed. The obtained 95$\%$- and respectively $68\%$-confidence intervals are $\{1\%, 45\%\}$ and $\{5\%, 28\%\}$.
\section{Discussion}
\label{sec:discussion}
In Section~\ref{sec:previousconstraints}, the occurrence constraints are compared to previous research which used surveys other than K2. In Section~\ref{sec:msoccurrence}, the WD occurrence is compared to the MS occurrence. In Section~\ref{sec:pmsevolution}, the results are discussed in context with PMS planetary system evolution. Lastly, in Section~\ref{sec:caveatsassumptions} we highlight the caveats and important assumptions of this work.
\subsection{Comparison with previous surveys}
\label{sec:previousconstraints}
The first attempt to constrain the non-disintegrating SBs occurrence around WDs used data from the Wide Angle Search for Planets (WASP) project \citep{Pollacco2006} to detect eclipsing SBs \citep{Faedi2011}. They used a smaller WD sample (194) and WASP has a lower sensitivity than K2. Consequently, strong constraints were derived only for short-period gas giants, which occur among ${<10} \%$ of all WDs. The results presented here provide stronger constraints for the full parameter space that was covered.\\
More recently, the Panoramic Survey Telescope And Rapid Response System (Pan-STARRS) data searched for eclipses around WDs \citep{Fulton2014}. They used a larger WD sample ($\sim$1718 WDs) and Pan-STARSS has equal or better detection probability than K2 for radii $\geq 2 R_{\oplus}$ for the whole range for which they derived constraints ($0.01-0.04 \ \rm{au}$). Accordingly, they constrained the maximum occurrence of hot Jupiters around WDs ($10-20 \ R_{\oplus}$ in between $0.01-0.04 \ \rm{au}$) to be $\leq 0.5 \%$, which is a stronger limit than what is presented here. On the other hand, K2 provides stronger constraints for radii $\leq 2 \ R_{\oplus}$, particularly for $R \leq 1.5 \ R_{\oplus}$, which are hardly covered by the ground-based Pan-STARRS survey. K2 most likely has stronger constraints than Pan-STARRS for $> 0.04 \ \rm{au}$ (outside of the range for which \cite{Fulton2014} derived constraints), but the low transit probabilities for WDs make it hard to probe this part of the parameter space even with K2.
\subsection{Comparison with MS planetary systems}
\label{sec:msoccurrence}
To compare the WD planetary system architecture to that of MS stars the limits in Figure~\ref{fig:occurrenceLCandSC} are divided into the more familiar MS classes: hot Jupiters, hot (super-)Earths and habitable-zone (super-)Earths. This is motivated by other studies which showed similarities between MS planetary systems and scaled-down versions of low-mass M-dwarfs planetary systems, the Jupiter-moon system \citep{Muirhead2012}, our solar system and exoplanetary systems around more massive stars \citep{Fulton2014}. Additionally, the classes of habitable-zone giants and sub-Earth-sized objects are added. The inner boundary of the CHZ is estimated as $0.005 \ \rm{au}$ ($\sim 4 \ \rm{h}$), 'hot' Jupiters and (super-)Earths are defined to orbit within this boundary. Altogether, the classification adopted in this work is summarized by Table~\ref{tab:classes}.\\
\begin{table}
\caption{The exoplanetary classification system for WDs used in this work. $\Delta a$ indicates the semi-major axis range, $\Delta P$ the period range and $\Delta R_p$ the planet radius range. The ranges are bases on scaled-down versions of the conventional MS classes. An average WD is implicitly assumed.}
\label{tab:classes}
\centering
\begin{tabular}{ l  c  c  r }
        \hline
        {Class} & {$\Delta a$} & {$\Delta P$} & {$\Delta R_p$} \\
    {} & {[au]} & [h] & [$R_{\earth}$] \\
        \hline
        Hot Jupiters & $<$0.005 & $<$4 & $>$10 \\
        Hot (super-)Earths & $<$0.005 & $<$4 & 0.5-2 \\
        Habitable zone (super-)Earths & 0.005-0.02 & 4-32 & 0.5-2 \\
        Habitable zone giants & 0.005-0.02 & 4-32 & $>$2 \\
        Sub-Earth-sized objects & - & - & $<0.5$ \\
    \hline
\end{tabular}
\end{table}
Our results constrain the occurrence of hot Jupiters $\leq 1.5 \% \ (0.6 \%)$, which implies hot Jupiters are rare or non-existing. This is supported by the strong constraints on hot Jupiters by WASP and Pan-STARRS \citep{Faedi2011, Fulton2014}, and similar or lower than the the occurrence of hot Jupiters around MS stars, which is estimated as 0.3-1.5$\%$ \citep[e.g.][]{Marcy2005, Gould2006, Cumming2008, Howard2010, Mayor2011, Wright2012}. Hot (super-)Earths of 1-2 $R_{\earth}$ occur among $\leq 9.8 \% \ (3.8 \%)$ of all WDs and for 0.5-1 $R_{\earth}$ among $\leq 18 \% \ (7.5 \%)$ of all WDs. The occurrence of MS close-in hot (super-)Earths is estimated as 23$\%$ \citep{Howard2010}, so that hot (super-)Earths around WDs are likely rarer than around MS stars.\\
Habitable zone (super-)Earths of 1-2 $R_{\earth}$ occur among $\leq 27.5 \% \ (11.5 \%)$ of all WDs and for 0.5-1 $R_{\earth}$ among $\leq 46.3 \% \ (21.1 \%)$. Therefore their occurrence is similar or less than their occurrence around MS stars of $\sim$22$\%$ \citep{Petigura2013}. Ice giants $2-4 \ R_{\earth}$ in the HZ are constrained to occur $\leq 17.0 \% \ (6.8 \%)$ and gas giants are constrained to occur $\leq 5.0 \% \ (1.9 \%)$.\\
Small transit probabilities, shorter transit durations and smaller transit depths make non-disintegrating SBs around MS stars $\leq 0.5 \ R_{\oplus}$ almost undetectable. In contrast to MS stars, K2 can constrain these objects around WDs. In the 'hot' -regime $0.5-1 \ R_{\earth}$ occur among $\leq 18.4 \% \ (7.5 \%)$ of all WDs and for $0.25-0.5 \ R_{\earth}$ occur $\leq 39 \% \ (17 \%)$. Even $0.125-0.25 \ R_{\earth}$ non-disintegrating SBs can be constrained to occur among $\leq 50 \% \ (23 \%)$ all WDs.
\subsection{Occurrence constraints in context with PMS planetary system evolution}
\label{sec:pmsevolution}
The inner planets close to the host star are dragged inwards during the AGB-phase due to tidal interactions and are destroyed \citep[see e.g.][]{doi:10.1093/mnras/stt137}. K2 can usefully constrain SBs up to $\sim 0.1 \ \rm{au}$. This is well within the region where both gas giants and terrestrial planets are theorized to be destroyed \citep{Mustill2012}. Unless another mechanism exists to migrate outer SBs inwards again, a null detection is thus in good agreement with the theory. Furthermore, the Roche radius for gas giants around WDs is estimated as $L_{\rm{R}} = 0.01 \ \rm{au}$ \citep{Fulton2014}. Since $L_R > 0.005 \ \rm{au}$, the outer boundary for the 'hot' regime, it seems unlikely the analogy of MS hot Jupiters around WDs are sustainable.\\
The outer planets are less effected by tides and expand their orbits due to the stellar mass loss during the AGB-phase \citep[see e.g.][]{Veras2016a, Veras2016b}. After the AGB-phase dynamical instabilities can perturb planets \citep{Veras2014} or exomoons \citep{Payne2016} inwards into the region $\leq 0.1 \ \rm{au}$ from the WD, the region covered by K2. However, if these SBs remain there just briefly, the probability of detecting one decreases dramatically. Simulation show liberated exomoons orbit within 0.1 au a fraction of $\sim 10^{-3}$ of the full simulation lifetime \citep[Figure 3]{Payne2016}. Rigorously assuming all WDs have liberated moons, their occurrence would be $\sim$0.1 $\%$, much lower than the limits derived here. The occurrence constraints for larger objects are stronger, but these objects are less likely to orbit close to the WD for a long time \citep[Figure 3]{Payne2016}. Nonetheless, these simulations did not take tidal circulization into account \citep{Veras2014, Payne2016}. Tidal circulization around WDs is theorized to occur similarly to MS stars \citep{Mustill2013}. This implies tidal interactions arise from the deformation of the planet rather than the WD. Both \cite{Veras2014} and \cite{Payne2016} neglected the interior structure of the planets, for it would complicate their simulations enormously. If tidal circulization successfully stabilizes perturbed planets it could greatly increase the prospects of detecting them.\\
The estimated disintegrating planet occurrence ($\sim 12 \%$ within $\{1\%, 45\% \}$ at 95$\%$-confidence) is most-likely smaller than the fraction of metal polluted WDs of 25-50$\%$ \citep{Cassan2012}. This suggests about half of the metal polluted WDs do not host a disintegrating planet. It remains difficult to interpret these results: the SBs fuelling metal pollution might have been fully disintegrated already, smaller undetectable asteroids may account for the metal pollution or debris disks may fuel the metal pollution instead. 
Nonetheless, we can conclude disintegrating objects around WDs are most-likely rarer than initially estimated by \cite{Vanderburg2015}, due to a larger observed sample with no new such objects detected.
\subsection{Caveats and important assumptions}
\label{sec:caveatsassumptions}
The results presented in Figures~\ref{fig:coverage},~\ref{fig:occurrenceLCandSC} and~\ref{fig:HZconstraints} are limited by both our data and necessary assumptions. We highlight the caveats and important assumptions of this work:
\begin{itemize}
\item The sample size: The size of a WD causes transit probabilities to be much lower for WDs than for MS stars. Therefore, the sample size is the main limiting factor for the occurrence constraints. With only a few more K2 campaigns remaining, the total number of WDs observed by the satellite will not increase dramatically. For the radii larger than $2 \ R_{\oplus}$ Kepler has a similar detection probability as Pan-STARRS \citep{Fulton2014}. However, since Pan-STARRS has a larger sample size, K2 is outperformed by Pan-STARRS for this part of the parameter space.
\item Transit probabilities: the probability for an eclipse rapidly decreases as a function of distance to the star. Therefore, periods $> 10 \ \rm{d}$ are poorly constrained, even though K2 has longer campaign durations of $\sim \ 80 \ \rm{d}$.
\item K2 data quality and sensitivity: the smallest SBs are poorly constrained due to low detection probability of K2. Nevertheless, as discussed in Section~\ref{sec:previousconstraints}, K2 has better detection probability than all currently operating surveys for radii $< 1.5 R_{\oplus}$, with some constraints available even for objects as small as $0.125 R_\oplus$.
\item WD parameters during the transit injection: a WD of mass $0.6 \ M_{\odot}$ and radius $0.012 \ R_{\odot}$ was assumed for all transit injections. The vast majority of WDs have parameters close to these, although a small fraction spans a wider range \citep{Tremblay2016}. 
\item WD contaminant stars: In many cases, WDs are faint stars (for the K2 sample, the median magnitude is $K_\mathrm{p} = 18.6$), which nearby bright stars may contaminate the light curves and lower the true sensitivity. To mitigate this, WDs with known dM companions were excluded, and pixel files were visually inspected for companion stars.
\item Disintegrating planet occurrence constraints: We assumed other disintegrating planets have similar transit signals as EPIC 201563164. However, follow-up observations show the transit depth is highly variable, most-likely due to events of extra dust production by irregular fragmentation of the parent body \cite{Gansicke2016, Gary2016}. Therefore, we are much more sensitive to disintegrating planets during their active stages compared to their quiescent stages. Additionally, it was assumed the detection efficiency equals that of a non-disintegrating body transit with a similar transit depth. However, their transit signal shapes differ, because of the large dust tail of a disintegrating planet. Therefore, we emphasize taking into account disintegrating planets varying activity levels and proper modelling and injection of their light curves and is most-likely necessary to derive more accurate constraints for disintegrating objects around WDs.
\end{itemize}
\section{Conclusions}
\label{sec:conclusions}
There is still a lot to learn about the ultimate fate of planetary systems and more observations are favourable. This research analysed 1148 high-probability WDs observed by K2. The WD images were converted into light curves using a new pipeline optimized for WDs observed by K2. A BLS detection algorithm was used to detect transiting/eclipsing objects around the WDs. Ten eclipsing objects were found: the one known disintegrating planet and nine likely stellar objects. From the null detection of new SBs and transit injections, upper limits on the occurrence of planets and SBs were calculated as a function of radius and orbital period. The primary conclusions of this work are:
\begin{itemize}
        \item{At short orbital periods ($< 40$~days), we can constrain the occurrence of small objects around WDs, outperforming previous constraints for asteroid-sized objects up to 1.5~$R_\oplus$.}
        \item{In line with theoretical predictions, hot Jupiters are rare or non-existing and occur among $< 1.5 \%$ of all WDs.} 
        \item{The occurrence of habitable Earth-sized planets ($1-2 \ R_{\oplus}$) around WDs is $< 28 \%$, approximately equal or less than their MS occurrence.}
        \item{The disintegrating planet occurrence is estimated as $\sim 12 \%$, which is lower than the estimated fraction of metal polluted WDs. However, detailed light curve modelling and taking varying activity levels into account is most-likely necessary to derive more accurate constraints.}
        \item{The data reduction and the transit/eclipse detection algorithm used in this work, specifically optimized for WDs observed by K2, successfully retrieved all of the previously known eclipsing objects blindly, adding confidence to the observed null detection of other SBs. The light curves are publicly available.}
\end{itemize}
Further constraining the occurrence of objects orbiting WDs would require larger samples, and any discovery of transiting planets or SBs would be of great interest, because the small size of WDs makes transits ideally suited for e.g.\ the search for exomoons or atmospheric follow-up studies. More transiting WDs may be detected with the Large Synoptic Survey Telescope \citep{LSST}, TESS \citep{ricker2014,raddi2017}, PLATO \citep{rauer2014}, NGTS \citep{wheatley2013}, and Evryscope \citep{law2015} .
\section*{Acknowledgements}
\thanks{
We are grateful to J.~J.~Hermes for providing his catalogue to crosscheck likely WDs, as well as helpful suggestions that improved this manuscript. We also thank Ignas Snellen for comments on the style and structure of this manuscript.}
\bibliographystyle{mnras}
\bibliography{mybibliography.bib} 

\begin{thebibliography}{}
\makeatletter
\relax
\def\mn@urlcharsother{\let\do\@makeother \do\$\do\&\do\#\do\^\do\_\do\%\do\~}
\def\mn@doi{\begingroup\mn@urlcharsother \@ifnextchar [ {\mn@doi@}
  {\mn@doi@[]}}
\def\mn@doi@[#1]#2{\def\@tempa{#1}\ifx\@tempa\@empty \href
  {http://dx.doi.org/#2} {doi:#2}\else \href {http://dx.doi.org/#2} {#1}\fi
  \endgroup}
\def\mn@eprint#1#2{\mn@eprint@#1:#2::\@nil}
\def\mn@eprint@arXiv#1{\href {http://arxiv.org/abs/#1} {{\tt arXiv:#1}}}
\def\mn@eprint@dblp#1{\href {http://dblp.uni-trier.de/rec/bibtex/#1.xml}
  {dblp:#1}}
\def\mn@eprint@#1:#2:#3:#4\@nil{\def\@tempa {#1}\def\@tempb {#2}\def\@tempc
  {#3}\ifx \@tempc \@empty \let \@tempc \@tempb \let \@tempb \@tempa \fi \ifx
  \@tempb \@empty \def\@tempb {arXiv}\fi \@ifundefined
  {mn@eprint@\@tempb}{\@tempb:\@tempc}{\expandafter \expandafter \csname
  mn@eprint@\@tempb\endcsname \expandafter{\@tempc}}}

\bibitem[\protect\citeauthoryear{{Adams}, {Jackson}  \& {Endl}}{{Adams}
  et~al.}{2016}]{Adams}
{Adams} E.~R.,  {Jackson} B.,   {Endl} M.,  2016, \mn@doi [Astrophysical
  Journal] {10.3847/0004-6256/152/2/47}, \href
  {http://adsabs.harvard.edu/abs/2016AJ....152...47A} {152, 47}

\bibitem[\protect\citeauthoryear{Agol}{Agol}{2011}]{Agol2011}
Agol E.,  2011, \mn@doi [The Astrophysical Journal]
  {10.1088/2041-8205/731/2/L31}, 731, L31

\bibitem[\protect\citeauthoryear{Althaus, C{\'{o}}rsico, Isern  \&
  Garc{\'{i}}a-Berro}{Althaus et~al.}{2010}]{Althaus2010}
Althaus L.~G.,  C{\'{o}}rsico A.~H.,  Isern J.,   Garc{\'{i}}a-Berro E.,  2010,
  \mn@doi [The Astronomy and Astrophysics Review, Volume 18, Issue 4,
  pp.471-566] {10.1007/s00159-010-0033-1}, 18, 471

\bibitem[\protect\citeauthoryear{Bonsor \& Veras}{Bonsor \&
  Veras}{2015}]{Bonsor2015}
Bonsor A.,  Veras D.,  2015, \mn@doi [Monthly Notices of the Royal Astronomical
  Society, Volume 454, Issue 1, p.53-63] {10.1093/mnras/stv1913}, 454, 53

\bibitem[\protect\citeauthoryear{Bonsor, Mustill  \& Wyatt}{Bonsor
  et~al.}{2011}]{Bonsor2011}
Bonsor A.,  Mustill A.,   Wyatt M.,  2011, \mn@doi [Monthly Notices of the
  Royal Astronomical Society, Volume 414, Issue 2, pp. 930-939.]
  {10.1111/j.1365-2966.2011.18524.x}, 414, 930

\bibitem[\protect\citeauthoryear{Burgasser, Kirkpatrick, Reid, Brown, Miskey
  \& Gizis}{Burgasser et~al.}{2002}]{Burgasser2002}
Burgasser A.~J.,  Kirkpatrick J.~D.,  Reid I.~N.,  Brown M.~E.,  Miskey C.~L.,
   Gizis J.~E.,  2002, \mn@doi [The Astrophysical Journal, Volume 586, Issue 1,
  pp. 512-526.] {10.1086/346263}, 586, 512

\bibitem[\protect\citeauthoryear{Cassan et~al.,}{Cassan
  et~al.}{2012}]{Cassan2012}
Cassan A.,  et~al., 2012, \mn@doi [Nature, Volume 481, Issue 7380, pp. 167-169
  (2012).] {10.1038/nature10684}, 481, 167

\bibitem[\protect\citeauthoryear{Cumming, Butler, Marcy, Vogt, Wright  \&
  Fischer}{Cumming et~al.}{2008}]{Cumming2008}
Cumming A.,  Butler R.~P.,  Marcy G.~W.,  Vogt S.~S.,  Wright J.~T.,   Fischer
  D.~A.,  2008, \mn@doi [Publications of the Astronomical Society of Pacific,
  Volume 120, Issue 867, pp. 531 (2008).] {10.1086/588487}, 120, 531

\bibitem[\protect\citeauthoryear{Debes, Kilic, Faedi, Shkolnik, Lopez-Morales,
  Weinberger, Slesnick  \& West}{Debes et~al.}{2012}]{Debes2012}
Debes J.~H.,  Kilic M.,  Faedi F.,  Shkolnik E.~L.,  Lopez-Morales M.,
  Weinberger A.~J.,  Slesnick C.,   West R.~G.,  2012, ]
  {10.1088/0004-637X/754/1/59}

\bibitem[\protect\citeauthoryear{Dong, Wang, Lin  \& Liu}{Dong
  et~al.}{2010}]{Dong2010}
Dong R.,  Wang Y.,  Lin D. N.~C.,   Liu X.~W.,  2010, \mn@doi [The
  Astrophysical Journal, Volume 715, Issue 2, pp. 1036-1049 (2010).]
  {10.1088/0004-637X/715/2/1036}, 715, 1036

\bibitem[\protect\citeauthoryear{Drake et~al.,}{Drake et~al.}{2010}]{Drake2010}
Drake A.~J.,  et~al., 2010

\bibitem[\protect\citeauthoryear{Faedi, West, Burleigh, Goad  \& Hebb}{Faedi
  et~al.}{2011}]{Faedi2011}
Faedi F.,  West R.~G.,  Burleigh M.~R.,  Goad M.~R.,   Hebb L.,  2011, \mn@doi
  [Monthly Notices of the Royal Astronomical Society]
  {10.1111/j.1365-2966.2010.17488.x}, 410, 899

\bibitem[\protect\citeauthoryear{Farihi, Zuckerman  \& Becklin}{Farihi
  et~al.}{2005}]{Farihi2005}
Farihi J.,  Zuckerman B.,   Becklin E.~E.,  2005, \mn@doi [The Astronomical
  Journal] {10.1086/491707}, 130, 2237

\bibitem[\protect\citeauthoryear{{Farihi}, {Barstow}, {Redfield}, {Dufour}  \&
  {Hambly}}{{Farihi} et~al.}{2010}]{2010MNRAS.404.2123F}
{Farihi} J.,  {Barstow} M.~A.,  {Redfield} S.,  {Dufour} P.,   {Hambly} N.~C.,
  2010, \mn@doi [Monthly Notices of the Royal Astronomical Society]
  {10.1111/j.1365-2966.2010.16426.x}, \href
  {http://adsabs.harvard.edu/abs/2010MNRAS.404.2123F} {404, 2123}

\bibitem[\protect\citeauthoryear{Fossati et~al.,}{Fossati
  et~al.}{2014}]{Fossati2014}
Fossati L.,  et~al., 2014, \mn@doi [Proceedings of the International
  Astronomical Union] {10.1017/S1743921315004998}, 10, 325

\bibitem[\protect\citeauthoryear{Frewen \& Hansen}{Frewen \&
  Hansen}{2014}]{Frewen2014}
Frewen S. F.~N.,  Hansen B. M.~S.,  2014, \mn@doi [Monthly Notices of the Royal
  Astronomical Society, Volume 439, Issue 3, p.2442-2458]
  {10.1093/mnras/stu097}, 439, 2442

\bibitem[\protect\citeauthoryear{Fulton et~al.,}{Fulton
  et~al.}{2014}]{Fulton2014}
Fulton B.~J.,  et~al., 2014, \mn@doi [The Astrophysical Journal, Volume 796,
  Issue 2, article id. 114, 9 pp. (2014).] {10.1088/0004-637x/796/2/114}, 796

\bibitem[\protect\citeauthoryear{G{\"{a}}nsicke et~al.,}{G{\"{a}}nsicke
  et~al.}{2016}]{Gansicke2016}
G{\"{a}}nsicke B.~T.,  et~al., 2016, ] {10.3847/2041-8205/818/1/L7}

\bibitem[\protect\citeauthoryear{Gary, Rappaport, Kaye, Alonso  \&
  Hambsch}{Gary et~al.}{2016}]{Gary2016}
Gary B.~L.,  Rappaport S.,  Kaye T.~G.,  Alonso R.,   Hambsch F.-J.,  2016,
  MNRAS, 000, 1

\bibitem[\protect\citeauthoryear{Girven, Brinkworth, Farihi, G{\"{a}}nsicke,
  Hoard, Marsh  \& Koester}{Girven et~al.}{2012}]{Girven2012}
Girven J.,  Brinkworth C.~S.,  Farihi J.,  G{\"{a}}nsicke B.~T.,  Hoard D.~W.,
  Marsh T.~R.,   Koester D.,  2012, \mn@doi [The Astrophysical Journal, Volume
  749, Issue 2, article id. 154, 11 pp. (2012).] {10.1088/0004-637X/749/2/154},
  749

\bibitem[\protect\citeauthoryear{Gould et~al.,}{Gould et~al.}{2006}]{Gould2006}
Gould A.,  et~al., 2006, \mn@doi [The Astrophysical Journal, Volume 644, Issue
  1, pp. L37-L40.] {10.1086/505421}, 644, L37

\bibitem[\protect\citeauthoryear{Heller, Homeier, Dreizler  \&
  {\O}stensen}{Heller et~al.}{2009}]{Heller2009}
Heller R.,  Homeier D.,  Dreizler S.,   {\O}stensen R.,  2009, \mn@doi [A{\&}A]
  {10.1051/0004-6361:200810632}, 496, 191

\bibitem[\protect\citeauthoryear{{Hermes}, {G{\"a}nsicke}, {Gentile Fusillo},
  {Raddi}, {Hollands}, {Dennihy}, {Fuchs}  \& {Redfield}}{{Hermes}
  et~al.}{2017}]{hermes2017}
{Hermes} J.~J.,  {G{\"a}nsicke} B.~T.,  {Gentile Fusillo} N.~P.,  {Raddi} R.,
  {Hollands} M.~A.,  {Dennihy} E.,  {Fuchs} J.~T.,   {Redfield} S.,  2017,
  \mn@doi [\mnras] {10.1093/mnras/stx567}, \href
  {http://adsabs.harvard.edu/abs/2017MNRAS.468.1946H} {468, 1946}

\bibitem[\protect\citeauthoryear{Howard et~al.,}{Howard
  et~al.}{2010}]{Howard2010}
Howard A.~W.,  et~al., 2010, \mn@doi [The Astrophysical Journal, Volume 721,
  Issue 2, pp. 1467-1481 (2010).] {10.1088/0004-637X/721/2/1467}, 721, 1467

\bibitem[\protect\citeauthoryear{{Howell} et~al.,}{{Howell}
  et~al.}{2014}]{Howell2014}
{Howell} S.~B.,  et~al., 2014, \mn@doi [\pasp] {10.1086/676406}, \href
  {http://adsabs.harvard.edu/abs/2014PASP..126..398H} {126, 398}

\bibitem[\protect\citeauthoryear{{Jura} \& {Young}}{{Jura} \&
  {Young}}{2014}]{2014AREPS..42...45J}
{Jura} M.,  {Young} E.~D.,  2014, \mn@doi [Annual Review of Earth and Planetary
  Sciences] {10.1146/annurev-earth-060313-054740}, \href
  {http://adsabs.harvard.edu/abs/2014AREPS..42...45J} {42, 45}

\bibitem[\protect\citeauthoryear{Kilic, Gould  \& Koester}{Kilic
  et~al.}{2009}]{Kilic2009}
Kilic M.,  Gould A.,   Koester D.,  2009, \mn@doi [The Astrophysical Journal,
  Volume 705, Issue 2, pp. 1219-1225 (2009).] {10.1088/0004-637X/705/2/1219},
  705, 1219

\bibitem[\protect\citeauthoryear{Kleinman et~al.,}{Kleinman
  et~al.}{2004}]{Kleinman2004}
Kleinman S.~J.,  et~al., 2004, \mn@doi [The Astrophysical Journal, Volume 607,
  Issue 1, pp. 426-444.] {10.1086/383464}, 607, 426

\bibitem[\protect\citeauthoryear{Koester, G{\"{a}}nsicke  \& Farihi}{Koester
  et~al.}{2014}]{Koester2014}
Koester D.,  G{\"{a}}nsicke B.~T.,   Farihi J.,  2014, \mn@doi [Astronomy {\&}
  Astrophysics, Volume 566, id.A34, 20 pp.] {10.1051/0004-6361/201423691}, 566

\bibitem[\protect\citeauthoryear{Kov{\'{a}}cs, Zucker  \& Mazeh}{Kov{\'{a}}cs
  et~al.}{2002}]{Kovacs2002}
Kov{\'{a}}cs G.,  Zucker S.,   Mazeh T.,  2002, \mn@doi [Astronomy and
  Astrophysics, v.391, p.369-377 (2002)] {10.1051/0004-6361:20020802}, 391, 369

\bibitem[\protect\citeauthoryear{Kunitomo, Ikoma, Sato, Katsuta  \&
  Ida}{Kunitomo et~al.}{2011}]{Kunitomo2011}
Kunitomo M.,  Ikoma M.,  Sato B.,  Katsuta Y.,   Ida S.,  2011, \mn@doi [The
  Astrophysical Journal, Volume 737, Issue 2, article id. 66, 8 pp. (2011).]
  {10.1088/0004-637X/737/2/66}, 737

\bibitem[\protect\citeauthoryear{{LSST Science Collaboration} et~al.,}{{LSST
  Science Collaboration} et~al.}{2009}]{LSST}
{LSST Science Collaboration} et~al., 2009, preprint, \href
  {http://adsabs.harvard.edu/abs/2009arXiv0912.0201L} {} (\mn@eprint {arXiv}
  {0912.0201})

\bibitem[\protect\citeauthoryear{{Law} et~al.,}{{Law} et~al.}{2015}]{law2015}
{Law} N.~M.,  et~al., 2015, \mn@doi [\pasp] {10.1086/680521}, \href
  {http://adsabs.harvard.edu/abs/2015PASP..127..234L} {127, 234}

\bibitem[\protect\citeauthoryear{Loeb \& Maoz}{Loeb \& Maoz}{2013}]{Loeb2013}
Loeb A.,  Maoz D.,  2013, ] {10.1093/mnrasl/slt026}

\bibitem[\protect\citeauthoryear{Lund, Handberg, Davies, Chaplin  \&
  Jones}{Lund et~al.}{2015}]{Lund2015}
Lund M.~N.,  Handberg R.,  Davies G.~R.,  Chaplin W.~J.,   Jones C.~D.,  2015,
  \mn@doi [The Astrophysical Journal, Volume 806, Issue 1, article id. 30, 15
  pp. (2015).] {10.1088/0004-637X/806/1/30}, 806

\bibitem[\protect\citeauthoryear{Mandel \& Agol}{Mandel \&
  Agol}{2002}]{Mandel2002}
Mandel K.,  Agol E.,  2002, \mn@doi [The Astrophysical Journal, Volume 580,
  Issue 2, pp. L171-L175.] {10.1086/345520}, 580, L171

\bibitem[\protect\citeauthoryear{Marcy, Butler, Fischer, Vogt, Wright, Tinney
  \& Jones}{Marcy et~al.}{2005}]{Marcy2005}
Marcy G.,  Butler R.~P.,  Fischer D.~A.,  Vogt S.~S.,  Wright J.~T.,  Tinney
  C.~G.,   Jones H. R.~A.,  2005, \mn@doi [Progress of Theoretical Physics
  Supplement, No. 158, pp. 24-42] {10.1143/PTPS.158.24}, 158, 24

\bibitem[\protect\citeauthoryear{Mayor et~al.,}{Mayor et~al.}{2011}]{Mayor2011}
Mayor M.,  et~al., 2011, eprint arXiv:1109.2497

\bibitem[\protect\citeauthoryear{McCarthy \& Zuckerman}{McCarthy \&
  Zuckerman}{2004}]{McCarthy2004}
McCarthy C.,  Zuckerman B.,  2004, \mn@doi [The Astronomical Journal]
  {10.1086/383559}, 127, 2871

\bibitem[\protect\citeauthoryear{McCree}{McCree}{1971}]{McCree1971}
McCree K.~J.,  1971, \mn@doi [Agricultural Meteorology]
  {10.1016/0002-1571(71)90022-7}, 9, 191

\bibitem[\protect\citeauthoryear{Muirhead et~al.,}{Muirhead
  et~al.}{2012}]{Muirhead2012}
Muirhead P.~S.,  et~al., 2012, \mn@doi [The Astrophysical Journal, Volume 747,
  Issue 2, article id. 144, 16 pp. (2012).] {10.1088/0004-637X/747/2/144}, 747

\bibitem[\protect\citeauthoryear{Mullally, Kilic, Reach, Kuchner, von Hippel,
  Burrows  \& Winget}{Mullally et~al.}{2006}]{Mullally2006}
Mullally F.,  Kilic M.,  Reach W.~T.,  Kuchner M.~J.,  von Hippel T.,  Burrows
  A.,   Winget D.~E.,  2006, ] {10.1086/511858}

\bibitem[\protect\citeauthoryear{Mullally, Winget, Degennaro, Jeffery,
  Thompson, Chandler  \& Kepler}{Mullally et~al.}{2008}]{Mullally2008}
Mullally F.,  Winget D.~E.,  Degennaro S.,  Jeffery E.,  Thompson S.~E.,
  Chandler D.,   Kepler S.~O.,  2008, Extreme Solar Systems, ASP Conference
  Series, Vol. 398, proceedings of the conference held 25-29 June, 2007, at
  Santorini Island, Greece. Edited by D. Fischer, F. A. Rasio, S. E. Thorsett,
  and A. Wolszczan, p.163, 398, 163

\bibitem[\protect\citeauthoryear{Mustill \& Villaver}{Mustill \&
  Villaver}{2012}]{Mustill2012}
Mustill A.~J.,  Villaver E.,  2012, \mn@doi [The Astrophysical Journal, Volume
  761, Issue 2, article id. 121, 13 pp. (2012).] {10.1088/0004-637X/761/2/121},
  761

\bibitem[\protect\citeauthoryear{Mustill, Veras  \& Villaver}{Mustill
  et~al.}{2013}]{Mustill2013}
Mustill A.~J.,  Veras D.,   Villaver E.,  2013, Mon. Not. R. Astron. Soc.XXXX)
  Printed, 000, 1

\bibitem[\protect\citeauthoryear{Nordhaus \& Spiegel}{Nordhaus \&
  Spiegel}{2012}]{Nordhaus2012}
Nordhaus J.,  Spiegel D.~S.,  2012, \mn@doi [Monthly Notices of the Royal
  Astronomical Society, Volume 432, Issue 1, p.500-505] {10.1093/mnras/stt569},
  432, 500

\bibitem[\protect\citeauthoryear{Parsons et~al.,}{Parsons
  et~al.}{2017}]{Parsons2017}
Parsons S.~G.,  et~al., 2017, \mn@doi [Monthly Notices of the Royal
  Astronomical Society, Volume 471, Issue 1, p.976-986]
  {10.1093/mnras/stx1610}, 471, 976

\bibitem[\protect\citeauthoryear{Payne, Veras, Gaensicke  \& Holman}{Payne
  et~al.}{2016}]{Payne2016}
Payne M.~J.,  Veras D.,  Gaensicke B.~T.,   Holman M.~J.,  2016, \mn@doi
  [Monthly Notices of the Royal Astronomical Society, Volume 464, Issue 3,
  p.2557-2564] {10.1093/mnras/stw2585}, 464, 2557

\bibitem[\protect\citeauthoryear{Petigura, Howard  \& Marcy}{Petigura
  et~al.}{2013}]{Petigura2013}
Petigura E.~A.,  Howard A.~W.,   Marcy G.~W.,  2013, \mn@doi [Proceedings of
  the National Academy of Sciences, vol. 110, issue 48, pp. 19273-19278]
  {10.1073/pnas.1319909110}, 110, 19273

\bibitem[\protect\citeauthoryear{Pollacco et~al.,}{Pollacco
  et~al.}{2006}]{Pollacco2006}
Pollacco D.~L.,  et~al., 2006, \mn@doi [The Publications of the Astronomical
  Society of the Pacific, Volume 118, Issue 848, pp. 1407-1418.]
  {10.1086/508556}, 118, 1407

\bibitem[\protect\citeauthoryear{{Raddi} et~al.,}{{Raddi}
  et~al.}{2017}]{raddi2017}
{Raddi} R.,  et~al., 2017, preprint, \href
  {http://adsabs.harvard.edu/abs/2017arXiv170809394R} {} (\mn@eprint {arXiv}
  {1708.09394})

\bibitem[\protect\citeauthoryear{Rappaport et~al.,}{Rappaport
  et~al.}{2017}]{Rappaport2017}
Rappaport S.,  et~al., 2017, MNRAS, 000, 1

\bibitem[\protect\citeauthoryear{{Rauer} et~al.,}{{Rauer}
  et~al.}{2014}]{rauer2014}
{Rauer} H.,  et~al., 2014, \mn@doi [Experimental Astronomy]
  {10.1007/s10686-014-9383-4}, \href
  {http://adsabs.harvard.edu/abs/2014ExA...tmp...41R} {}

\bibitem[\protect\citeauthoryear{Rebassa-Mansergas, Gaensicke, Schreiber,
  Koester  \& Rodriguez-Gil}{Rebassa-Mansergas
  et~al.}{2009}]{Rebassa-Mansergas2009}
Rebassa-Mansergas A.,  Gaensicke B.~T.,  Schreiber M.~R.,  Koester D.,
  Rodriguez-Gil P.,  2009, \mn@doi [Monthly Notices of the Royal Astronomical
  Society, Volume 402, Issue 1, pp. 620-640.]
  {10.1111/j.1365-2966.2009.15915.x}, 402, 620

\bibitem[\protect\citeauthoryear{{Ricker} et~al.,}{{Ricker}
  et~al.}{2014}]{ricker2014}
{Ricker} G.~R.,  et~al., 2014, in Society of Photo-Optical Instrumentation
  Engineers (SPIE) Conference Series. p.~20 (\mn@eprint {arXiv} {1406.0151}),
  \mn@doi{10.1117/12.2063489}

\bibitem[\protect\citeauthoryear{Sanchis-Ojeda et~al.,}{Sanchis-Ojeda
  et~al.}{2015}]{Sanchis-Ojeda2015}
Sanchis-Ojeda R.,  et~al., 2015, \mn@doi [The Astrophysical Journal]
  {10.1088/0004-637X/812/2/112}, 812, 112

\bibitem[\protect\citeauthoryear{Sandhaus, Debes, Ely, Hines  \&
  Bourque}{Sandhaus et~al.}{2016}]{Sandhaus2016}
Sandhaus P.~H.,  Debes J.~H.,  Ely J.,  Hines D.~C.,   Bourque M.,  2016, ]
  {10.3847/0004-637X/823/1/49}

\bibitem[\protect\citeauthoryear{Silvestri et~al.,}{Silvestri
  et~al.}{2006}]{Silvestri2006}
Silvestri N.~M.,  et~al., 2006, \mn@doi [The Astronomical Journal]
  {10.1086/499494}, 131, 1674

\bibitem[\protect\citeauthoryear{Silvotti et~al.,}{Silvotti
  et~al.}{2012}]{Silvotti2012}
Silvotti R.,  et~al., 2012, \mn@doi [Monthly Notices of the Royal Astronomical
  Society] {10.1111/j.1365-2966.2012.21232.x}, 424, 1752

\bibitem[\protect\citeauthoryear{{Tremblay}, {Cummings}, {Kalirai},
  {G{\"a}nsicke}, {Gentile-Fusillo}  \& {Raddi}}{{Tremblay}
  et~al.}{2016}]{Tremblay2016}
{Tremblay} P.-E.,  {Cummings} J.,  {Kalirai} J.~S.,  {G{\"a}nsicke} B.~T.,
  {Gentile-Fusillo} N.,   {Raddi} R.,  2016, \mn@doi [\mnras]
  {10.1093/mnras/stw1447}, \href
  {http://adsabs.harvard.edu/abs/2016MNRAS.461.2100T} {461, 2100}

\bibitem[\protect\citeauthoryear{{Van Eylen} et~al.,}{{Van Eylen}
  et~al.}{2016}]{Vaneylen2016}
{Van Eylen} V.,  et~al., 2016, \mn@doi [\apj] {10.3847/0004-637X/820/1/56},
  \href {http://adsabs.harvard.edu/abs/2016ApJ...820...56V} {820, 56}

\bibitem[\protect\citeauthoryear{Vanderburg et~al.,}{Vanderburg
  et~al.}{2015}]{Vanderburg2015}
Vanderburg A.,  et~al., 2015, \mn@doi [Nature] {10.1038/nature15527}, 526, 546

\bibitem[\protect\citeauthoryear{Veras \& Gaensicke}{Veras \&
  Gaensicke}{2014}]{Veras2014}
Veras D.,  Gaensicke B.~T.,  2014, \mn@doi [Monthly Notices of the Royal
  Astronomical Society, Volume 447, Issue 2, p.1049-1058]
  {10.1093/mnras/stu2475}, 447, 1049

\bibitem[\protect\citeauthoryear{Veras, Mustill, G{\"{a}}nsicke, Redfield,
  Georgakarakos, Bowler  \& Lloyd}{Veras et~al.}{2016a}]{Veras2016a}
Veras D.,  Mustill A.~J.,  G{\"{a}}nsicke B.~T.,  Redfield S.,  Georgakarakos
  N.,  Bowler A.~B.,   Lloyd M. J.~S.,  2016a, Mon. Not. R. Astron. Soc, 000, 1

\bibitem[\protect\citeauthoryear{Veras, Mustill, G{\"{a}}nsicke, Redfield,
  Georgakarakos, Bowler  \& Lloyd}{Veras et~al.}{2016b}]{Veras2016b}
Veras D.,  Mustill A.~J.,  G{\"{a}}nsicke B.~T.,  Redfield S.,  Georgakarakos
  N.,  Bowler A.~B.,   Lloyd M. J.~S.,  2016b, Mon. Not. R. Astron. Soc, 000, 1

\bibitem[\protect\citeauthoryear{Villaver \& Livio}{Villaver \&
  Livio}{2009}]{Villaver2009}
Villaver E.,  Livio M.,  2009, \mn@doi [The Astrophysical Journal Letters,
  Volume 705, Issue 1, pp. L81-L85 (2009).] {10.1088/0004-637X/705/1/L81}, 705,
  L81

\bibitem[\protect\citeauthoryear{Villaver, Livio, Mustill  \& Siess}{Villaver
  et~al.}{2014}]{Villaver2014}
Villaver E.,  Livio M.,  Mustill A.~J.,   Siess L.,  2014, \mn@doi [The
  Astrophysical Journal, Volume 794, Issue 1, article id. 3, 15 pp. (2014).]
  {10.1088/0004-637X/794/1/3}, 794

\bibitem[\protect\citeauthoryear{Voyatzis, Hadjidemetriou, Veras  \&
  Varvoglis}{Voyatzis et~al.}{2013}]{doi:10.1093/mnras/stt137}
Voyatzis G.,  Hadjidemetriou J.~D.,  Veras D.,   Varvoglis H.,  2013, \mn@doi
  [Monthly Notices of the Royal Astronomical Society] {10.1093/mnras/stt137},
  430, 3383

\bibitem[\protect\citeauthoryear{{Wheatley} et~al.,}{{Wheatley}
  et~al.}{2013}]{wheatley2013}
{Wheatley} P.~J.,  et~al., 2013, in European Physical Journal Web of
  Conferences. p. 13002 (\mn@eprint {arXiv} {1302.6592}),
  \mn@doi{10.1051/epjconf/20134713002}

\bibitem[\protect\citeauthoryear{Wright, Marcy, Howard, Johnson, Morton  \&
  Fischer}{Wright et~al.}{2012}]{Wright2012}
Wright J.~T.,  Marcy G.~W.,  Howard A.~W.,  Johnson J.~A.,  Morton T.~D.,
  Fischer D.~A.,  2012, \mn@doi [The Astrophysical Journal]
  {10.1088/0004-637X/753/2/160}, 753

\bibitem[\protect\citeauthoryear{Wyatt, Farihi, Pringle  \& Bonsor}{Wyatt
  et~al.}{2014}]{Wyatt2014}
Wyatt M.~C.,  Farihi J.,  Pringle J.~E.,   Bonsor A.,  2014, ]
  {10.1093/mnras/stu183}

\bibitem[\protect\citeauthoryear{Xu \& Jura}{Xu \& Jura}{2011}]{Xu2011}
Xu S.,  Jura M.,  2011, \mn@doi [The Astrophysical Journal, Volume 745, Issue
  1, article id. 88, 14 pp. (2012).] {10.1088/0004-637X/745/1/88}, 745

\bibitem[\protect\citeauthoryear{Xu, Ertel, Wahhaj, Milli, Scicluna  \&
  Bertrang}{Xu et~al.}{2015}]{Xu2015}
Xu S.,  Ertel S.,  Wahhaj Z.,  Milli J.,  Scicluna P.,   Bertrang G. H.~M.,
  2015, \mn@doi [Astronomy {\&} Astrophysics, Volume 579, id.L8, 5 pp.]
  {10.1051/0004-6361/201526179}, 579

\bibitem[\protect\citeauthoryear{Zuckerman, Koester, Reid  \& Hunsch}{Zuckerman
  et~al.}{2003}]{Zuckerman2003}
Zuckerman B.,  Koester D.,  Reid I.~N.,   Hunsch M.,  2003, \mn@doi [The
  Astrophysical Journal] {10.1086/377492}, 596, 477

\bibitem[\protect\citeauthoryear{Zuckerman, Melis, Klein, Koester  \&
  Jura}{Zuckerman et~al.}{2010}]{Zuckerman2010}
Zuckerman B.,  Melis C.,  Klein B.,  Koester D.,   Jura M.,  2010, \mn@doi [The
  Astrophysical Journal, Volume 722, Issue 1, pp. 725-736 (2010).]
  {10.1088/0004-637X/722/1/725}, 722, 725

\makeatother
\end{thebibliography}
\bsp    
\label{lastpage}
\end{document}